\documentclass[10pt,final,journal,twocolumn]{IEEEtran}

\usepackage{amsfonts}
\usepackage{cite}
\usepackage{amsmath}
\usepackage{amssymb}
\usepackage{graphicx}
\usepackage{subfigure}
\usepackage{color}
\newcommand{\tabincell}[2]{\begin{tabular}{@{}#1@{}}#2\end{tabular}}

\title{Transceiver Design for Multi-user Multi-\\antenna Two-way Relay Cellular Systems}

\author{Can~Sun,~\IEEEmembership{Student Member,~IEEE,} Chenyang~Yang,~\IEEEmembership{Senior Member,~IEEE,}
Yonghui~Li,~\IEEEmembership{Senior Member,~IEEE,}
and~Branka~Vucetic,~\IEEEmembership{Fellow,~IEEE}
\thanks{Manuscript received May 30, 2011; revised February 16, 2012
and May 3, 2012.}
\thanks {This work was supported in part by the
International S$\&$T Cooperation Program of China (ISCP) under Grant
No. 2008DFA12100, and in part by Australian Research Council (ARC)
Discovery Projects DP120100190, DP110103324, DP0877090 and Linkage
Projects LP0991663, and has been presented in part at the IEEE
Global Telecommunications Conference (GLOBECOM), Miami, USA,
December 2010, and at the IEEE International Conference on
Acoustics, Speech, and Signal Processing (ICASSP), Prague, Czech
Republic, May 2011.}
\thanks{C. Sun and C. Yang are with the School of Electronics and
Information Engineering, Beihang University, Beijing 100191, China
(email: saga@ee.buaa.edu.cn, cyyang@buaa.edu.cn).}
\thanks{Y. Li and B. Vucetic are with the School of Electrical and Information
Engineering, University of Sydney, NSW 2006, Australia (email:
\{lyh,branka\}@ee.usyd.edu.au).}}

\begin{document}
\maketitle

\begin{abstract}
In this paper, we design interference free transceivers for
multi-user two-way relay systems, where a multi-antenna base station
(BS) simultaneously exchanges information with multiple
single-antenna users via a multi-antenna amplify-and-forward relay
station (RS). To offer a performance benchmark and provide useful
insight into the transceiver structure, we employ alternating
optimization to find optimal transceivers at the BS and RS that
maximizes the bidirectional sum rate. We then propose a low
complexity scheme, where the BS transceiver is the zero-forcing
precoder and detector, and the RS transceiver is designed to balance
the uplink and downlink sum rates. Simulation results demonstrate
that the proposed scheme is superior to the existing zero forcing
and signal alignment schemes, and the performance gap between the
proposed scheme and the alternating optimization is minor.
\end{abstract}


\section{Introduction}
Two-way relay (TWR) techniques have attracted considerable interest
owing to its high spectral efficiency. Most of prior works study TWR
systems with single user pair, where two users exchange information
via a single relay station (RS) \cite{Louie2010,Zhang2009,Oech2009}.
Various transmission schemes have been proposed for single antenna
nodes \cite{Louie2010} and multi-antenna nodes
\cite{Zhang2009,Oech2009}.

Recently, the design for TWR systems is extended to multi-user cases
\cite{Chen2009,Joung2010,CY2010,Jit2009,Esli2008a,Toh2009,DK2011,Yang2008a},
which can be roughly divided into two categories based on the system
topologies, i.e., symmetric and asymmetric systems. In symmetric
systems \cite{Chen2009,Joung2010,CY2010}, multiple user pairs
exchange information via a RS. In asymmetric systems, a base station
(BS) exchanges messages with multiple users
\cite{Jit2009,Esli2008a,Yang2008a,Toh2009,DK2011}, which is a
typical scenario of cellular networks.

In this paper, we study multi-user TWR cellular system, where a
multi-antenna BS communicates with multiple single-antenna users
bidirectionally via a multi-antenna amplify-and-forward (AF) relay.
Owing to the importance from practical perspective, there is a
considerable amount of work on designing transceivers for such a
system \cite{Esli2008a,Yang2008a,Toh2009,DK2011}. However, its
transceiver optimization is challenging due to the complicated
interference among multiple users in the broadcast and multi-access
phases, and even its bidirectional sum capacity is still not
available until now.

Allocating orthogonal time or frequency resources to the uplink and
downlink signals of different users is an immediate way to eliminate
the interference \cite{Jit2009}, with which existing single-user TWR
techniques can be directly applied. Since this is far from optimal,
a further attempt is to introduce an interference free constraint,
which is essentially the zero-forcing (ZF) principle. Though also
suboptimal in a sense of sum rate, such a design can capture the
inherent degrees of freedom of the system, which is an approximate
characterization of the capacity at the high signal-to-noise (SNR)
level. Along this line, several ZF-principle based transceivers have
been proposed. Considering that the RS is equipped with multiple
antennas, a natural solution is to apply ZF transceiver at the RS to
separate all the signals from and to the BS and users
\cite{Esli2008a}. This ZF scheme employs orthogonal spatial
resources to differentiate different links, thereby the RS should be
equipped with enough antennas. To remove all the interference, at
least $2N$ antennas are required at the RS for a system with $N$
antennas at the BS and $N$ single antenna users. When the RS is only
with $N$ antennas, the multiple antennas at the BS also need to be
exploited to ensure interference free transmission. In
\cite{Toh2009,DK2011,Yang2008a}, the concept of signal alignment
(SA) \cite{Lee2010} is employed to reduce the number of interference
experienced at the relay. The SA scheme exploits the
self-interference cancelation (SIC) \cite{Rankov07} ability of TWR.
Its basic idea is to project the uplink and downlink signals of each
user onto the same spatial direction at the RS through proper BS
precoding, such that the RS can separate $N$ superimposed signals.
After receiving a superimposed signal forwarded by the RS, each user
removes its transmitted uplink signal via SIC, and obtains its
desired downlink signal.

Both the ZF and SA schemes are based on ZF-principle. Nonetheless,
they are not the only interference free solution\footnote{By using
the terminology ``interference free solutions'', we refer to the
transmit strategies that can remove all interference. These
solutions include the ZF beamforming and ZF detector, the SA scheme,
as well as the transmit schemes using orthogonal frequency or time
resources, which can null the interference thoroughly.}. In fact, by
analyzing the feasibility of interference free constraints for
multi-user multi-antenna TWR cellular systems, it is not hard to
show that the SA scheme is the unique solution only for special
antenna configurations, and the ZF scheme ensures interference free
transmission only when the number of antennas at the RS is
sufficiently large. Moreover, both of them are designed as low
complexity schemes without taking into account the sum rate.

In this paper, we strive to find a low complexity interference free
transceiver towards maximizing sum rate under general antenna
settings. To provide a performance benchmark as well as useful
insight into the transceiver structure, we employ a standard
alternating optimization technique \cite{Hath2002} to optimize the
BS and RS transceivers aiming at maximizing bidirectional sum rate
under interference free constraints. In order to develop a low
complexity transceiver scheme, we fix the BS transceiver as the
optimal BS precoder and detector in high power region found from the
alternating optimization. Based on which we first optimize the RS
transceiver to separately maximize the uplink and downlink sum rates
and then balance the uplink and downlink sum rates to maximize the
bidirectional sum rate. Simulation results show that the balanced
scheme performs very close to the alternating optimization solution,
and outperforms existing ZF and SA schemes under various scenarios.

The rest of the paper is organized as follows. Section
\ref{sec:system_model} describes the system model. Section
\ref{sec:optimizing} introduces the alternating optimization
solution. The balanced transceiver scheme is proposed in
\ref{sec:heuristic}. Simulation results are given in section
\ref{sec:simulation}, and conclusions are drawn in section
\ref{sec:conclusion}. The major symbols used in the paper are
summarized in Table~\ref{ta:parameter_list}.

\begin{table}[htp]
\renewcommand{\arraystretch}{1.3}
\caption{List of important symbols} \label{ta:parameter_list}
\begin{center}
\begin{tabular}{c|p{5.8cm}}
\hline \hline
$N_B$, $N_R$, $N_U$  & BS or RS antenna number or user number\\
\hline
$\mathbf{H}_{br}, \mathbf{H}_{ur}$ & Channel matrix from the BS or from all users to the RS \\
\hline
$\mathbf{h}_{ir}$ & Channel vector from the $i$th user to the RS\\
\hline
$\overline{\mathbf{H}}_{ir}$  & Channel matrix from all users other than the $i$th user to the RS.\\
                               & It is obtained from $\mathbf{H}_{ur}$ with the $i$th column, $\mathbf{h}_{ir}$,
                                being removed.\\
\hline
$\mathbf{W}_{bt}$, $\mathbf{W}_{br}$ & BS transmit or receive weighting matrix\\
\hline
$\mathbf{w}_{bti}$, $\mathbf{w}_{bri}$ & The $i$th column of $\mathbf{W}_{bt}$ or $\mathbf{W}_{br}$\\
\hline
$\mathbf{W}_{r}$ & RS weighting matrix\\
\hline
$\mathbf{x}_b$ & Downlink signal vector transmitted by the BS\\
\hline
$\mathbf{x}_u$ & Uplink signal vector transmitted by all users\\
\hline
$\mathbf{y}_r$ & RS's received signal vector in first phase\\
\hline
$\mathbf{y}_b, y_{ui}$ & BS's or the $i$th user's received signal in second phase\\
\hline
$P_B$, $P_R$, $P_U$& The transmit power of BS or RS or a single user \\
\hline
$N_0$ & Noise variance\\
\hline
$R_U$, $R_D$, $R_S$ & Uplink or downlink or bidirectional sum rate\\
\hline
$\mathbf{I}_N$ & Identity matrix of size $N$\\
\hline \hline
$(\cdot)^T$, $(\cdot)^H$, $(\cdot)^*$ & Transpose, conjugate transpose or conjugate of a matrix\\
\hline
$\lVert\cdot\rVert$, $(\cdot)^{\dagger}$ & Norm or pseudo inverse of a matrix\\
\hline
$S_\bot(\mathbf{X})$ & Orthogonal subspace of matrix $\mathbf{X}$\\
                    & $S_\bot(\mathbf{X}) = \mathbf{I}-\mathbf{X}^H(\mathbf{X}\mathbf{X}^H)^{-1}\mathbf{X}$ if $\mathbf{X}$ is a wide matrix\\
                    & $S_\bot(\mathbf{X}) = \mathbf{I}-\mathbf{X}(\mathbf{X}^H\mathbf{X})^{-1}\mathbf{X}^H$ if $\mathbf{X}$ is a high matrix\\
\hline
$\text{diag}(\mathbf{m})$ & Diagonal matrix whose diagonal elements are the elements of vector $\mathbf{m}$ \\
\hline
$E(\cdot)$ & Mean value of a random variable\\
\hline
\end{tabular}
\end{center}
\end{table}

\section{System Model}
\label{sec:system_model} 
\begin{figure}[htp]
\centering
\includegraphics[width=3.4in]{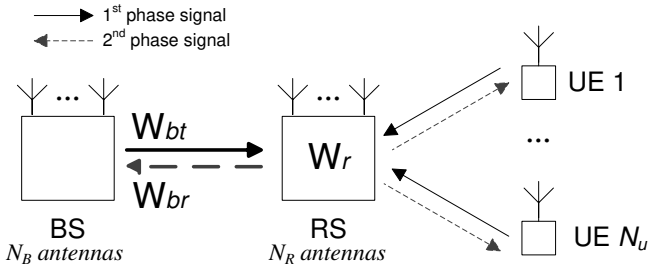}
\caption{System model of the multi-user multi-antenna TWR cellular
system}\label{fig:system_model}
\end{figure}

We consider a multi-user multi-antenna TWR
system, which consists of a BS equipped with $N_B$ antennas, a RS
equipped with $N_R$ antennas and $N_U$ single-antenna users. The BS
and multiple users exchange downlink and uplink information via the
RS, as shown in Fig.~\ref{fig:system_model}. The bidirectional
transmission takes place in two phases.

At the first phase, both the BS and multiple users transmit to the
RS. The received signal at the RS is given by
\begin{align}
\label{eq:1st phase relay received} \mathbf{y}_r & =
\mathbf{H}_{br}\mathbf{W}_{bt}\mathbf{x}_b +
\sqrt{P_U}\mathbf{H}_{ur}\mathbf{x}_u + \mathbf{n}_r,
\end{align}
where $\mathbf{H}_{br}\in\mathbb{C}^{N_R\times N_B}$ is the channel
matrix from the BS to the RS, $\mathbf{H}_{ur} = (\mathbf{h}_{1r},
\cdots, \mathbf{h}_{N_Ur})$,
$\mathbf{h}_{ir}\in\mathbb{C}^{N_R\times 1}$ is the channel vector
from the $i$th user to the RS, $\mathbf{x}_b$ and $\mathbf{x}_u$ are
the downlink and uplink signal vectors to and from $N_U$ users and
we assume $E(\mathbf{x}_b\mathbf{x}_b^H)
=E(\mathbf{x}_u\mathbf{x}_u^H)=\mathbf{I}_{N_U}$, $P_U$ is the
transmit power of each user, $\mathbf{n}_r$ is the Gaussian noise
vector at the RS with zero mean and covariance matrix
$N_0\mathbf{I}_{N_R}$, and $\mathbf{W}_{bt}\in\mathbb{C}^{N_B\times
N_U}$ is the precoder matrix at the BS, which satisfies the transmit
power constraint as follows
\begin{align}
\label{eq:BS Power Constraint} \lVert\mathbf{W}_{bt} \rVert^2 \leq
P_B,
\end{align}
where $P_B$ is the maximal transmit power of the BS.

At the second phase, the RS precodes its received signals and then
broadcasts them to the BS and users. The received signals at the BS
and the $i$th user are respectively given by
\begin{align}
\label{eq:2nd phase BS received}
\mathbf{y}_b & = \mathbf{W}_{br}^T(\mathbf{H}_{br}^T\mathbf{W}_r\mathbf{y}_r + \mathbf{n}_b),\\
\label{eq:2nd phase user received} y_{ui} & =
\mathbf{h}_{ir}^T\mathbf{W}_r\mathbf{y}_r + n_{ui},~~(1\leq i\leq
N_U),
\end{align}
where $\mathbf{W}_r\in\mathbb{C}^{N_R\times N_R}$ is the weighting
matrix at the RS, $\mathbf{W}_{br}\in\mathbb{C}^{N_B\times N_U}$ is
the receive weighting matrix at the BS, and $\mathbf{n}_b$ and
$n_{ui}$ are Gaussian noises at the BS and the $i$th user, each with
zero mean and variance $N_0$.

The RS weighting matrix should satisfy the transmit power constraint
$E(\lVert \mathbf{W}_r\mathbf{y}_r \rVert^2) \leq P_R$, which can be
rewritten as follows after substituting (\ref{eq:1st phase relay
received}),
\begin{align}
\label{eq:RS Power Contraint}
\lVert\mathbf{W}_r\mathbf{H}_{br}\mathbf{W}_{bt} \rVert^2 +
P_U\lVert\mathbf{W}_r\mathbf{H}_{ur} \rVert^2  +
N_0\lVert\mathbf{W}_r \rVert^2\leq P_R,
\end{align}
where $P_R$ is the maximal transmit power of the RS\footnote{We do
not consider power control at the BS and RS. The inequality power
constraints are for simplifying the optimization.}.

All channels are assumed independent quasi-static flat fading and we
consider time division duplexing for simplicity, hence the channels
in the 1st and 2nd phases are assumed reciprocal. We assume that the
BS and RS have global channel information of all links as in
\cite{Yang2008a,Toh2009,DK2011}.

\section{Transceiver Design Based on Alternating Optimization}
\label{sec:optimizing} Even after we introduce the interference free
constraints, the problem of jointly optimizing BS and RS
transceivers that maximizes the bidirectional sum rate of multi-user
multi-antenna TWR systems is still non-convex and is very hard to
deal with. In this section, we employ a standard tool, alternating
optimization \cite{Hath2002}, to solve the optimization problem,
which can serve as a performance benchmark for the interference free
transceivers.

Substituting (\ref{eq:1st phase relay received}) into (\ref{eq:2nd
phase BS received}), the received signal at the BS can be rewritten
as
\begin{align}
\label{eq:BS received at last} \mathbf{y}_b = &
\mathbf{W}_{br}^T\mathbf{H}_{br}^T\mathbf{W}_r\mathbf{H}_{br}\mathbf{W}_{bt}\mathbf{x}_b+
\sqrt{P_U}\mathbf{W}_{br}^T\mathbf{H}_{br}^T\mathbf{W}_r\mathbf{H}_{ur}\mathbf{x}_u+\nonumber\\
& \mathbf{W}_{br}^T\mathbf{H}_{br}^T\mathbf{W}_r\mathbf{n}_r +
\mathbf{W}_{br}^T\mathbf{n}_b,
\end{align}
where the first term is the transmitted signal of the BS in the
first phase which can be removed by SIC, the second term is the
desired uplink signal, and the last two terms are the noise
amplified by the RS and the noise at the BS receiver, respectively.

To eliminate the interference among the uplink signals, the
following constraint should be satisfied,
\begin{align}
\label{eq:constraint 1 for precoder}
\mathbf{w}_{bri}^T\mathbf{H}_{br}^T\mathbf{W}_r\mathbf{h}_{jr} = 0,
& ~~i \not=j,
\end{align}
where $\mathbf{w}_{bri}$ is the $i$th column of $\mathbf{W}_{br}$.

Substituting (\ref{eq:1st phase relay received}) into (\ref{eq:2nd
phase user received}), the received signals at the $i$th user can be
rewritten as
\begin{align}
\label{eq:MS received at last} y_{ui} = &
\mathbf{h}_{ir}^T\mathbf{W}_r\mathbf{H}_{br}\mathbf{W}_{bt}\mathbf{x}_b+
\sqrt{P_U}\mathbf{h}_{ir}^T\mathbf{W}_r\mathbf{H}_{ur}\mathbf{x}_u+
\mathbf{h}_{ir}^T\mathbf{W}_r\mathbf{n}_r\nonumber\\ & +
n_{ui}~~~(1\leq i\leq N_U),
\end{align}
where the first term consists of the downlink signals for all $N_U$
users, the second term consists of the transmitted signals from
$N_U$ users in the first phase, and the last two terms are noises.

To remove the interference, the BS and RS transceivers should
satisfy the following constraints,
\begin{align}
\label{eq:constraint 2 for precoder}
\mathbf{h}_{ir}^T\mathbf{W}_r\mathbf{H}_{br}\mathbf{w}_{btj} = 0, & ~~i \not=j,\\
\label{eq:constraint 3 for precoder}
\mathbf{h}_{ir}^T\mathbf{W}_r\mathbf{h}_{jr} = 0, & ~~i \not=j,
\end{align}
where $\mathbf{w}_{btj}$ is the $j$th column of $\mathbf{W}_{bt}$.

Considering the inter-user interference (IUI) free constraints
(\ref{eq:constraint 1 for precoder}), (\ref{eq:constraint 2 for
precoder}) and (\ref{eq:constraint 3 for precoder}) and the fact
that the self-interference can be canceled \cite{Rankov07}, the
receive SNR of the $i$th uplink and downlink signal can be
respectively obtained as,
\begin{align}
\label{eq:SINR} & SNR_{Ui} =
\frac{P_U\lvert\mathbf{w}_{bri}^T\mathbf{H}_{br}^T\mathbf{W}_r\mathbf{h}_{ir}\rvert^2}
{N_0\lVert\mathbf{w}_{bri}^T\mathbf{H}_{br}^T\mathbf{W}_r\rVert^2+N_0\lVert\mathbf{w}_{bri}^T
\rVert^2},\nonumber\\
& SNR_{Di} =
\frac{\lvert\mathbf{h}_{ir}^T\mathbf{W}_r\mathbf{H}_{br}\mathbf{w}_{bti}\rvert^2}
{N_0\lVert\mathbf{h}_{ir}^T\mathbf{W}_r\rVert^2+N_0}.
\end{align}
Then the bidirectional sum rate of the TWR system is\footnote{The
received non-white noise after amplifying and forwarding is treated
as white noise as in existing literature. This is in fact the worst
case of the problem, therefore the data rate obtained by
$\log_2(1+SNR)$ can serve as a lower bound.},
\begin{align}
\label{eq:sum rate}  R_S & = R_U + R_D =
\sum_{i=1}^{N_U}(R_{Ui}+R_{Di})\nonumber\\ & =
\sum_{i=1}^{N_U}\big[\frac{1}{2}\log_2(1+SNR_{Ui}) +
\frac{1}{2}\log_2(1+SNR_{Di})\big],
\end{align}
where $R_U$ and $R_D$ denote the uplink and downlink sum rate,
$R_{Ui}$ and $R_{Di}$ are the uplink and downlink data rates of the
$i$th user, and the pre-log factor $1/2$ is due to the half-duplex
constraint.

In the following, we optimize one of the three transceiver matrices
by fixing the other two.

\subsection{Optimization of Weighting Matrix $\mathbf{W}_r$ of RS}
Here we fix $\mathbf{W}_{bt}$ and $\mathbf{W}_{br}$, and optimize
$\mathbf{W}_r$ to maximize the bidirectional sum rate under the RS
transmit power constraint and the IUI-free constraints by solving
the following problem,
\begin{subequations}
\label{eq:optimization subproWr}
\begin{align}
\label{eq:optimization subproWr_obj} \underset{\mathbf{W}_r}{\max}
~~& R_S\\
\label{eq:optimization subproWr_st1}\text{s.t.~~} &
\text{(\ref{eq:RS Power Contraint}), (\ref{eq:constraint 1 for
precoder}), (\ref{eq:constraint 2 for precoder}) and
(\ref{eq:constraint 3 for precoder})}.
\end{align}
\end{subequations}

The bidirectional sum rate $R_S$ is not a convex function of
$\mathbf{W}_r$. To solve this non-convex problem and find the
maximum $R_S$, we employ the concept of \emph{rate profile}, which
is introduced in \cite{Moh2006} to characterize the boundary
rate-tuples of a capacity region. We introduce a vector $\pmb{\beta}
= (\beta_1,\cdots,\beta_{2N_U})$ to specify the rate profile, where
$\sum_{i=1}^{2N_U}\beta_i = 1$ and $\beta_i \geq 0$. Then by solving
the following optimization problem,
\begin{subequations}
\label{eq:Wr_maximin}
\begin{align}
\label{eq:Wr_maximin_obj} \underset{\mathbf{W}_r}{\max} ~~& R_S\\
\label{eq:Wr_maximin_st1} \text{s.t.~~} & R_{Ui} \geq
\beta_iR_S,~R_{Di} \geq \beta_{i+N_U}R_S,~1\leq
i\leq N_U,\\
\label{eq:Wr_maximin_st3} &\text{(\ref{eq:RS Power Contraint}),
(\ref{eq:constraint 1 for precoder}), (\ref{eq:constraint 2 for
precoder}) and (\ref{eq:constraint 3 for precoder}),}
\end{align}
\end{subequations}
we will achieve a boundary point of the achievable rate region
specified by each vector $\pmb{\beta}$. After searching the optimal
$\pmb{\beta}$ from all its possible values, we can find the optimal
boundary point corresponding to the maximum sum rate. For multi-user
case, it is too complicated to search all possible $\pmb{\beta}$. To
reduce the complexity, we use bisection algorithm \cite{Burden} to
search the optimal $\pmb{\beta}$ as in \cite{Esli2008a}. Although it
is hard to rigorously prove that the achievable rate region boundary
is a convex hull in terms of $\pmb{\beta}$, simulation results show
that bisection algorithm offers the same result as that of using
brute-force searching.

To solve the problem (\ref{eq:Wr_maximin}), we apply a similar
approach as in \cite{Zhang2009} to convert the optimization problem
(\ref{eq:Wr_maximin}) to a semidefinite programming (SDP) problem
with a rank-1 constraint, and then we resort to the widely used
semidefinite relaxation \cite{Luo2006} to handle the problem.

\subsection{Optimization of Transmit Weighting Matrix $\mathbf{W}_{bt}$ of BS}
\label{sec:Wbt_opt} In this subsection, we design $\mathbf{W}_{bt}$
given $\mathbf{W}_{br}$ and $\mathbf{W}_{r}$. Since the transmit
weighting matrix of the BS only affects downlink rate when
$\mathbf{W}_{br}$ and $\mathbf{W}_{r}$ are fixed, we design it to
maximize the downlink sum rate $R_D$. The design of
$\mathbf{W}_{bt}$ should consider the IUI free constraint
(\ref{eq:constraint 2 for precoder}) and the BS transmit power
constraint (\ref{eq:BS Power Constraint}). It is also associated
with the RS transmit power constraint (\ref{eq:RS Power Contraint}).
Then the optimization problem can be formulated as
\begin{subequations}
\label{eq:optimization subproWbt}
\begin{align}
\label{eq:optimization subproWbt_obj}
\underset{\mathbf{W}_{bt}}{\max~}
& R_D \\
\label{eq:optimization subproWbt_st1}\text{s.t.~}& \text{(\ref{eq:BS
Power Constraint}), (\ref{eq:RS Power Contraint}) and
(\ref{eq:constraint 2 for precoder})}.
\end{align}
\end{subequations}

This is also a non-convex problem, which can be solved by the same
method as that we used to solve problem (\ref{eq:optimization
subproWr}). Define a vector $\pmb{\beta}=
[\beta_1,\cdots,\beta_{N_U}]$, where $\sum_{i=1}^{N_U}\beta_i = 1$
and $\beta_i \geq 0$. The solution of (\ref{eq:optimization
subproWbt}) can be found from solving the following problem by
searching the optimal $\pmb{\beta}$,
\begin{subequations}
\label{eq:wbt_maximin}
\begin{align}
\label{eq:wbt_maximin_obj} \underset{\mathbf{W}_{bt}}{\max~}
& R_D\\
\label{eq:wbt_maximin_st1} \text{s.t.~}&R_{Di} \geq
\beta_iR_D,~~1\leq i\leq
N_U,\\
&\text{(\ref{eq:BS Power Constraint}), (\ref{eq:RS Power Contraint})
and (\ref{eq:constraint 2 for precoder})}.
\end{align}
\end{subequations}

Each of the BS and RS power constraints (\ref{eq:BS Power
Constraint}) and (\ref{eq:RS Power Contraint}) imposes a constraint
on the norm of a linear function of $\mathbf{W}_{bt}$. The IUI free
constraint (\ref{eq:constraint 2 for precoder}) is a linear
constraint on $\mathbf{W}_{bt}$. According to \cite{Luo2006}, the
rate tuple constraint (\ref{eq:wbt_maximin_st1}) can be converted to
linear constraints on $\mathbf{W}_{bt}$. Therefore the constraints
in (\ref{eq:wbt_maximin}) form a second-order-cone feasible region
\cite{Luo2006}, and the size of the feasible region depends on
$R_D$. Consequently, we can solve (\ref{eq:wbt_maximin}) by
searching the maximal $R_D$ that guarantees a non-empty feasible
region. Bisection method is applied to search $R_D$. We use the CVX
tool\cite{cvx} to check whether the feasible region is empty or not.
If it is not empty, the CVX tool will return a value of
$\mathbf{W}_{bt}$ in the feasible region. Finally, we will obtain
both the maximum value of $R_D$ and the optimal $\mathbf{W}_{bt}$.

\subsection{Optimization of Receive Weighting Matrix $\mathbf{W}_{br}$ of BS}
\label{sec:Wbr_opt} Given $\mathbf{W}_{bt}$ and $\mathbf{W}_{r}$,
$\mathbf{W}_{br}$ only affects uplink sum rate. Among the three
IUI-free constraints, $\mathbf{W}_{br}$ is only associate with
(\ref{eq:constraint 1 for precoder}). Therefore, the optimization
problem can be formulated as
\begin{align}
\label{eq:optimization subproWbr} \underset{\mathbf{W}_{br}}{\max}
~~& R_U\cr \text{s.t.~~} &
\mathbf{w}_{bri}^T\mathbf{H}_{br}^T\mathbf{W}_r\mathbf{h}_{jr} = 0,
~~i \not=j.
\end{align}

According to (\ref{eq:SINR}) and (\ref{eq:sum rate}), the data rate
of each uplink stream, $R_{Ui}$, is only a function of
$\mathbf{w}_{bri}$. Therefore, this problem can be decoupled into
$N_U$ subproblems. Since $R_{Ui}$ is a monotonic increasing function
of $SNR_{Ui}$, each subproblem can be formulated as
\begin{subequations}
\label{eq:optimization subprowbri}
\begin{align}
\label{eq:optimization subprowbri_obj}
\underset{\mathbf{w}_{bri}}{\max} ~~& SNR_{Ui}\\
\label{eq:optimization subprowbri_st} \text{s.t.~~} &
\mathbf{w}_{bri}^T\mathbf{H}_{br}^T\mathbf{W}_r\mathbf{h}_{jr} = 0,
~~j \not=i.
\end{align}
\end{subequations}

Any feasible $\mathbf{w}_{bri}$ should satisfy
$\mathbf{w}_{bri}^T\mathbf{H}_{br}^T\mathbf{W}_r\overline{\mathbf{H}}_{ir}
= \mathbf{0}$, where $\overline{\mathbf{H}}_{ir}$ is obtained from
channel matrix $\mathbf{H}_{ur}$ with the $i$th column being
removed. Define $\mathbf{U}_{ir}^{\bot}$ as a matrix consisting of
all the singular vectors of
$\mathbf{H}_{br}^T\mathbf{W}_r\overline{\mathbf{H}}_{ir}$
corresponding to its zero singular values. Then we have
\begin{align}
\label{eq:wbri_ortho} \mathbf{w}_{bri} & =
\mathbf{U}_{ir}^{\bot}\mathbf{x},
\end{align}
where $\mathbf{x}$ is an arbitrary vector.

Rewrite the expression of $SNR_{Ui}$ in (\ref{eq:SINR}) as follows,
\begin{align}
\label{eq:rayleigh_ratio} SNR_{Ui} = &
\frac{\mathbf{w}_{bri}^T(P_U\mathbf{H}_{br}^T\mathbf{W}_r\mathbf{h}_{ir}
\mathbf{h}_{ir}^H\mathbf{W}_r^H\mathbf{H}_{br}^*)\mathbf{w}_{bri}^*}
{\mathbf{w}_{bri}^T(N_0
\mathbf{H}_{br}^T\mathbf{W}_r\mathbf{W}_r^H\mathbf{H}_{br}^*+N_0\mathbf{I}_{N_B})\mathbf{w}_{bri}^*}\nonumber\\
\triangleq &
\frac{\mathbf{w}_{bri}^T\mathbf{K}_S\mathbf{w}_{bri}^*}{\mathbf{w}_{bri}^T\mathbf{K}_{IN}\mathbf{w}_{bri}^*}.
\end{align}
where $\mathbf{K}_S \triangleq
P_U\mathbf{H}_{br}^T\mathbf{W}_r\mathbf{h}_{ir}
\mathbf{h}_{ir}^H\mathbf{W}_r^H\mathbf{H}_{br}^*$ and
$\mathbf{K}_{IN} \triangleq N_0
\mathbf{H}_{br}^T\mathbf{W}_r\mathbf{W}_r^H\mathbf{H}_{br}^*+N_0\mathbf{I}_{N_B}$.

By substituting (\ref{eq:wbri_ortho}) and (\ref{eq:rayleigh_ratio}),
the optimization problem (\ref{eq:optimization subprowbri}) becomes
\begin{align}
\label{eq:optimization rayleighratio} \underset{\mathbf{x}}{\max}
~~& \frac{\mathbf{x}^T\mathbf{U}_{ir}^{\bot
T}\mathbf{K}_S\mathbf{U}_{ir}^{\bot
*}\mathbf{x}^*}{\mathbf{x}^T\mathbf{U}_{ir}^{\bot T}\mathbf{K}_{IN}\mathbf{U}_{ir}^{\bot
*}\mathbf{x}^*},
\end{align}
which is a generalized Rayleigh ratio problem. The optimal
$\mathbf{x}$ is the eigenvector of $\mathbf{U}_{ir}^{\bot
T}\mathbf{K}_S\mathbf{U}_{ir}^{\bot
*}(\mathbf{U}_{ir}^{\bot T}\mathbf{K}_{IN}\mathbf{U}_{ir}^{\bot
*})^{-1}$
corresponding to its largest eigenvalue \cite{VanTree}.

By now, we have solved the three problems (\ref{eq:optimization
subproWr}), (\ref{eq:optimization subproWbt}) and
(\ref{eq:optimization subproWbr}). When we find the alternating
optimization solution, we need to assign initial values for the
transceiver matrices, which should satisfy all the IUI free
constraints and the transmit power constraints. The initial values
are set according to the following procedure.

First, constraint (\ref{eq:constraint 3 for precoder}) can be
rewritten as a group of linear equations of $\mathbf{W}_r$ as $
(\mathbf{h}_{jr}^T\otimes\mathbf{h}_{ir}^T)\text{vec}(\mathbf{W}_r)
= 0, i \not=j$, where $\otimes$ denotes Kronecker product, and
$\text{vec}(\cdot)$ is the vectorization of a matrix by stacking its
columns. The general solution of this equation is given by
\begin{align}\label{ini}
\text{vec}(\mathbf{W}_r) = \mathbf{S}_\bot(\mathbf{K}_O)\mathbf{x},
\end{align}
where $\mathbf{K}_O$ is the matrix by stacking all
$\mathbf{h}_{ir}^T\otimes\mathbf{h}_{jr}^T$, $i\not = j$,
$\mathbf{S}_\bot(\cdot)$ is the orthogonal subspace of a matrix, and
$\mathbf{x}$ is an arbitrary vector.

We pick one $\mathbf{W}_r$ from the general solution. Then we
substitute the chosen $\mathbf{W}_r$ into (\ref{eq:constraint 1 for
precoder}) and (\ref{eq:constraint 2 for precoder}), find general
solutions of these two set of equations similar to (\ref{ini}), and
pick one $\mathbf{W}_{bt}$ and one $\mathbf{W}_{br}$ among the
general solutions. Finally, we multiply $\mathbf{W}_{bt}$ and
$\mathbf{W}_r$ with proper scalars to satisfy the BS and RS power
constraints.

After assigning the initial values, we alternately optimize one of
the three transceiver matrices by fixing the other two. The sum rate
must increase with each iteration, otherwise, the iteration is
terminated. Due to this requirement, the alternating procedure will
surely converge. Because of the non-convex nature of the
optimization problem, the converged solution is not guaranteed to be
globally optimal, and depends on the initial values. Nevertheless,
we can increase the probability to achieve the maximal bidirectional
sum rate by repeating the alternating optimization procedure with
multiple random initial values then picking the best solution.

\section{Balanced Transceivers}
\label{sec:heuristic} In this section, we design a low complexity
transceiver toward achieving maximal bidirectional sum rate under
the interference free constraints. To this end, we decouple the
joint optimization of the BS and RS transceivers resorting to the
asymptotic analysis in high power region. Specifically, we first
find the BS precoder and detector from analyzing asymptotic results
of the alternating optimization solution. Then we optimize the RS
transceiver based on the given BS transceiver, also in high power
region.
\subsection{BS Transceivers}
\subsubsection{BS Precoder} To obtain a closed
form solution, we consider an asymptotic region where the transmit
power of RS goes to infinity. When $P_R \to\infty$, the RS transmit
power constraint can be ignored, then the BS precoder optimization
problem in (\ref{eq:optimization subproWbt}) can be reformulated as
\begin{align}
\underset{\mathbf{W}_{bt}}{\max~} &
\sum_{i=1}^{N_U}\log_2(1+\frac{\lvert\mathbf{h}_{ir}^T\mathbf{W}_r\mathbf{H}_{br}\mathbf{w}_{bti}\rvert^2}
{N_0\lVert\mathbf{h}_{ir}^T\mathbf{W}_r\rVert^2+N_0}) \cr
\text{s.t.~}& \lVert\mathbf{W}_{bt} \rVert^2 \leq P_B,\cr
&\mathbf{h}_{ir}^T\mathbf{W}_r\mathbf{H}_{br}\mathbf{w}_{btj} = 0,
~i \not=j,
\end{align}
which can be viewed as linear precoder optimization for downlink
multi-user multi-antenna system with a channel matrix
$\mathbf{H}_{ur}^T\mathbf{W}_r\mathbf{H}_{br}$ that maximizes the
sum rate under interference free constraints and total transmit
power constraint. According to \cite{Wiesel2008}, the optimal
precoder is a ZF precoder with proper power allocation, i.e.,
\begin{align}
\label{eq:Wbt_heuristic} \mathbf{W}_{bt} =
(\mathbf{H}_{ur}^T\mathbf{W}_r\mathbf{H}_{br})^\dagger\mathbf{G}_b,
\end{align}
where $\mathbf{G}_b$ is a diagonal power allocation matrix. For
simplicity, we consider equal power allocation at the BS, i.e.,
\begin{align}
\label{eq:equal power BS} \lVert\mathbf{w}_{bti}\rVert^2 = P_B/N_U.
\end{align}
\subsubsection{BS Detector}
To obtain a closed form detector, we consider another asymptotic
region where the transmit power of the BS or users approaches
infinity. When $P_U\to \infty$ or $P_B\to\infty$, the received SNR
at the RS in the first phase goes to infinity, then the RS forwarded
noise can be neglected\footnote{This is not true for the case of
deep fading where the channel coefficient is approximately zero, but
such a case is of low probability.}. In this case, $\mathbf{K}_{IN}$
in (\ref{eq:optimization rayleighratio}) is $N_0\mathbf{I}_{N_B}$.
By solving the problem (\ref{eq:optimization rayleighratio}) and
applying (\ref{eq:wbri_ortho}), the optimal BS receiver vector
$\mathbf{w}_{bri}$ can be obtained as
$\mathbf{U}_{ir}^\bot\mathbf{U}_{ir}^{\bot
H}(\mathbf{H}_{br}^T\mathbf{W}_r\mathbf{h}_{ir})^*$, where
$\mathbf{U}_{ir}^\bot\mathbf{U}_{ir}^{\bot H}$ spans the orthogonal
subspace of
$\mathbf{H}_{br}^T\mathbf{W}_r\overline{\mathbf{H}}_{ir}$
\cite{VanTree}. Therefore, the optimal $\mathbf{w}_{bri}$ is the
projection of $\mathbf{H}_{br}^T\mathbf{W}_r\mathbf{h}_{ir}$ onto
the orthogonal subspace of
$\mathbf{H}_{br}^T\mathbf{W}_r\overline{\mathbf{H}}_{ir}$, i.e., the
optimal solution is a ZF receiver for the equivalent uplink channel
$\mathbf{H}_{br}^T\mathbf{W}_r\mathbf{H}_{ur}$, i.e.,
\begin{align}
\label{eq:Wbr_heuristic} \mathbf{W}_{br} =
[(\mathbf{H}_{br}^T\mathbf{W}_r\mathbf{H}_{ur})^\dagger]^T.
\end{align}

Note that the obtained BS precoder and detector in
(\ref{eq:Wbt_heuristic}) and (\ref{eq:Wbr_heuristic}) are not
optimal for practical systems with finite transmit power.
Nonetheless, later we will show by simulations that these ZF
transceivers perform fairly well even when the transmit powers are
finite.

\subsection{RS Transceiver}
Now we find the solution of RS transceiver from
(\ref{eq:optimization subproWr}) given the BS transceivers
(\ref{eq:Wbt_heuristic}) and (\ref{eq:Wbr_heuristic}). The IUI free
constraints (\ref{eq:constraint 1 for precoder}) and
(\ref{eq:constraint 2 for precoder}) are satisfied owing to the
usage of ZF transceivers at the BS, and thus can be removed. Note
that we consider equal power allocation in the BS precoder, then the
optimization problem of RS transceiver can be reformulated as
\begin{subequations}
\label{eq:heuristic_Wr_pro1}
\begin{align}
\label{eq:heuristic_Wr_pro1_obj}
\underset{\mathbf{W}_r}{\max~~}&R_S\\
\label{eq:heuristic_Wr_pro1_st1} \text{s.t.~~} &\text{(\ref{eq:RS
Power Contraint}), (\ref{eq:constraint 3 for precoder}),
(\ref{eq:Wbt_heuristic}), (\ref{eq:Wbr_heuristic}) and
(\ref{eq:equal power BS})}.
\end{align}
\end{subequations}

To find a low complexity solution for this non-convex problem, we
decouple it into two subproblems, which respectively maximize the
uplink and downlink sum rate. Then we combine these two solutions to
maximize the bidirectional sum rate.

When the transmit power of each user goes to zero, i.e., $P_U \to
0$\footnote{This is not conflict with the optimality conditions of
the ZF transceivers at the BS, which are $P_R\to \infty$ and either
$P_B$ or $P_U\to \infty$,}, the system uplink sum rate will approach
to zero, then $R_S \to R_D$. From (\ref{eq:SINR}) and (\ref{eq:sum
rate}), the downlink sum rate $R_D$ does not depend on
$\mathbf{W}_{br}$, therefore the constraint (\ref{eq:Wbr_heuristic})
in problem (\ref{eq:heuristic_Wr_pro1}) can be removed. Moreover, in
this case the RS received signal at the first phase $\mathbf{y}_r
\to \mathbf{H}_{br}\mathbf{W}_{bt}\mathbf{x}_b +\mathbf{n}_r$. Then
the RS power constraint (\ref{eq:RS Power Contraint}) can be
rewritten as
$\lVert\mathbf{W}_r\mathbf{H}_{br}\mathbf{W}_{bt}\rVert^2 +
N_0\lVert\mathbf{W}_r\rVert^2 \leq P_R$. Consequently, the problem
(\ref{eq:heuristic_Wr_pro1}) reduces to the following problem that
maximizes the downlink sum rate,
\begin{align}
\label{eq:Wr_D problem} \underset{\mathbf{W}_r}{\max~~}&R_D\cr
\text{s.t.~~}& \text{(\ref{eq:constraint 3 for precoder}),
(\ref{eq:Wbt_heuristic}), (\ref{eq:equal power BS}) and }\nonumber\\
& \lVert\mathbf{W}_r\mathbf{H}_{br}\mathbf{W}_{bt}\rVert^2 +
N_0\lVert\mathbf{W}_r\rVert^2 \leq P_R.
\end{align}

Similarly, when the BS transmit power goes to zero, i.e., $P_B \to
0$, the problem (\ref{eq:heuristic_Wr_pro1}) reduces to the
following problem that maximizes the uplink sum rate,
\begin{align}
\label{eq:Wr_U problem} \underset{\mathbf{W}_r}{\max~~}&R_U\cr
\text{s.t.~~} &\text{(\ref{eq:constraint 3 for precoder}),
(\ref{eq:Wbr_heuristic}) and }\nonumber\\
& P_U\lVert\mathbf{W}_r\mathbf{H}_{ur}\rVert^2 +
N_0\lVert\mathbf{W}_r\rVert^2 \leq P_R.
\end{align}

We will first solve these two subproblems, then combine the two
solutions of $\mathbf{W}_r$ to balance the uplink and downlink
rates, so as to maximize the bidirectional sum rate.

\subsubsection{Design of $\mathbf{W}_r$ From Subproblem (\ref{eq:Wr_D
problem})}\label{ssec:Wr_D}

We can show that the optimal solution of (\ref{eq:Wr_D problem}) has
the following structure (see Appendix),
\begin{align}
\label{eq:Wr_D structure} \mathbf{W}_r =
(\mathbf{H}_{ur}^T)^{\dag}\mathbf{G}_{r1}\mathbf{U}^T,
\end{align}
where $\mathbf{G}_{r1}$ is a diagonal matrix and each column of
$\mathbf{U}$ has unit norm, i.e., $\lVert \mathbf{u}_j\rVert^2 = 1$.

\begin{figure}[htp]
\centering
\includegraphics[width=3.4in]{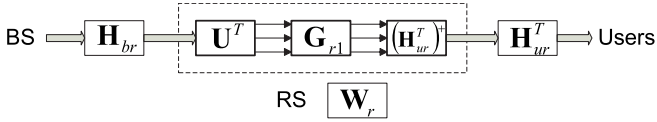}
\caption{Structure of the optimal RS transceiver for downlink
transmission}\label{fig:WrD_structure}
\end{figure}

The optimal structure of $\mathbf{W}_r$ can be intuitively explained
as follows. When $P_U \to 0$, there is only downlink transmission,
i.e., the RS receives signals from the BS and then forwards it to
the users. In this case, $(\mathbf{H}_{ur}^T)^{\dag}$ represents the
ZF precoder at the RS to broadcast signals to the users,
$\mathbf{G}_{r1}$ is a power allocation matrix for different signal
streams, and $\mathbf{U}^T$ is the receive weighting matrix at the
RS, which separates the $N_U$ downlink signals from the BS, see
Fig.~\ref{fig:WrD_structure}.

To obtain the power allocation matrix $\mathbf{G}_{r1} =
\text{diag}(p_{r1},\cdots,p_{rN_U})$, we simply let the
amplification coefficients at the RS for all streams to be
identical. Denote each column of $(\mathbf{H}_{ur}^T)^{\dag}$ as
$\mathbf{q}_i$. Since each downlink data stream is received by
$\mathbf{u}_i^T$, amplified by $p_{ri}$, and forwarded by
$\mathbf{q}_i$, we design $p_{ri}$ to ensure that each
$p_{ri}\mathbf{q}_i\mathbf{u}_i^T$ has the same norm.

Our next task is to design the receive weighting matrix
$\mathbf{U}$. Upon substituting (\ref{eq:Wr_D structure}), the
optimization problem (\ref{eq:Wr_D problem}) can be rewritten
as\\
\begin{subequations}
\label{eq:heursi_U_pro}
\begin{align}
\label{eq:heursi_U_pro_obj} \underset{\mathbf{U}}{\max~} &
\frac{1}{2}\sum_{i=1}^{N_U}\log_2\big(1+\frac{p_{ri}^2\lvert
\mathbf{u}_i^T\mathbf{H}_{br}\mathbf{w}_{bti}\rvert^2}
{N_0p_{ri}^2+N_0}\big)\\
\label{eq:heursi_U_pro_st1}\text{s.t.~} &
\mathbf{u}_i^T\mathbf{h}_{jr} = 0, ~i \not=j, ~~\lVert
\mathbf{u}_i\rVert = 1,\\
\label{eq:heursi_U_pro_st2} & \mathbf{W}_{bt} =
(\mathbf{U}^T\mathbf{H}_{br})^{\dagger}\mathbf{G}_{r1}^{-1}\mathbf{G}_b,\\
\label{eq:heursi_U_pro_st3}
& \lVert\mathbf{w}_{bti}\rVert^2 = P_B/N_U,\\
\label{eq:heursi_U_pro_st4}
&\lVert(\mathbf{H}_{ur}^T)^{\dag}\mathbf{G}_{r1}\mathbf{U}^T\mathbf{H}_{br}\mathbf{W}_{bt}\rVert^2+
N_0\lVert(\mathbf{H}_{ur}^T)^{\dag}\mathbf{G}_{r1}\mathbf{U}^T\rVert^2\nonumber\\
&\qquad\qquad\qquad\qquad\qquad\qquad\qquad\qquad\leq P_R.
\end{align}
\end{subequations}

This problem is non-convex, thereby we turn to find its suboptimal
solution. Since $\mathbf{U}^T$ acts as the receiver at the RS in
downlink transmission, we design it to maximize the ``data rate'' of
BS-RS transmission instead of the two-phase downlink
transmission\footnote{Since the RS does not decode message in AF
protocol, in fact there is no ``BS-RS transmission data rate''. We
use this terminology here for simplifying the optimization
problem.}.

During the BS-RS transmission, the half-duplex RS only receives
signals from the BS. Therefore, we do not consider the RS transmit
power constraint (\ref{eq:heursi_U_pro_st4}), which can be met later
by adjusting $\mathbf{G}_{r1}$. Then the BS-RS transmission rate
maximization problem is formulated as
\begin{subequations}
\label{eq:heursi_U_sub}
\begin{align}
\label{eq:heursi_U_sub_obj} \underset{\mathbf{U}}{\max~} &
\sum_{i=1}^{N_U}\log_2(1+\lvert
\mathbf{u}_i^T\mathbf{H}_{br}\mathbf{w}_{bti}\rvert^2/N_0)\\
\label{eq:heursi_U_sub_st1} \text{s.t.~} &
\text{(\ref{eq:heursi_U_pro_st1}), (\ref{eq:heursi_U_pro_st2}) and
(\ref{eq:heursi_U_pro_st3}).}
\end{align}
\end{subequations}

\emph{Remark 1:} If $P_R \to \infty$, the objective function
(\ref{eq:heursi_U_pro_obj}) will be the same as
(\ref{eq:heursi_U_sub_obj}) except for the pre-log factor $1/2$, and
the RS power constraint (\ref{eq:heursi_U_pro_st4}) can be omitted.
This means that the two optimization problems
 are approximately equivalent
when the RS has high transmit power.

Constraint (\ref{eq:heursi_U_pro_st2}) shows that $\mathbf{W}_{bt}$
is a pseudo inverse of $\mathbf{U}^T\mathbf{H}_{br}$ with power
allocation. Define $\overline{\mathbf{U}}_i$ as the matrix
$\mathbf{U}$ with the $i$th column $\mathbf{u}_i$ being removed.
Then using the principle of orthogonal projection\cite{VanTree}, we
obtain that
\begin{align*}
&\lvert\mathbf{u}_i^T\mathbf{H}_{br}\mathbf{w}_{bti}\rvert/\lVert\mathbf{w}_{bti}\rVert\nonumber\\
=&\lVert\mathbf{u}_i^T\mathbf{H}_{br}(\mathbf{I}-\mathbf{H}_{br}^H\overline{\mathbf{U}}_i^*(\overline{\mathbf{U}}_i^T\mathbf{H}_{br}\mathbf{H}_{br}^H\overline{\mathbf{U}}_i^*)^{-1}\overline{\mathbf{U}}_i^T\mathbf{H}_{br})
\rVert\nonumber\\ \triangleq &
\lVert\mathbf{u}_i^T\mathbf{H}_{br}\mathbf{S}_{\bot}(\overline{\mathbf{U}}_i^T\mathbf{H}_{br})\rVert.
\end{align*}
Substituting this expression and
(\ref{eq:heursi_U_pro_st3}) into (\ref{eq:heursi_U_sub_obj}), then
the problem (\ref{eq:heursi_U_sub}) can be rewritten as
\begin{align}
\label{eq:heursi_U_sub2} \underset{\mathbf{U}}{\max~} &
\sum_{i=1}^{N_U}\log_2(1+\frac{P_B}{N_UN_0}\lVert\mathbf{u}_i^T\mathbf{H}_{br}\mathbf{S}_{\perp}(\overline{\mathbf{U}}_i^T\mathbf{H}_{br})\rVert^2)\nonumber\\
\text{s.t.~} & \mathbf{u}_i^T\mathbf{h}_{jr} = 0, ~i \not=j,
~~\lVert \mathbf{u}_i\rVert = 1.
\end{align}
Constraints (\ref{eq:heursi_U_pro_st2}) and
(\ref{eq:heursi_U_pro_st3}) are omitted since the objective function
does not rely on $\mathbf{W}_{bt}$ now.

Solving problem (\ref{eq:heursi_U_sub2}) is nontrivial because we
need to jointly design all $\mathbf{u}_i$. To obtain a
low-complexity solution, we employ alternating optimization
\cite{Hath2002} again. We first initialize $\mathbf{U} =
\mathbf{0}$. Then we alternately optimize each of the $N_U$ columns
of $\mathbf{U}$. In each step, we optimize the $i$th column
$\mathbf{u}_i$ by solving the problem (\ref{eq:heursi_U_sub2}) with
all other columns $\overline{\mathbf{U}}_i$ being fixed. After each
step, we renew the matrix $\mathbf{U}$ by replacing its $i$th column
by the optimized $\mathbf{u}_i$. The procedure stops when the value
of objective function in (\ref{eq:heursi_U_sub2}) does not increase
any more. Simulations show that the procedure always converges after
each of the $N_U$ columns has been optimized once.

In the above procedure, we need to solve the optimization problem
(\ref{eq:heursi_U_sub2}) with fixed $\overline{\mathbf{U}}_i$. Note
that the constraint $\mathbf{u}_i^T\mathbf{h}_{jr} = 0$, $i\not=j$
can be rewritten as $\mathbf{u}_i^T\overline{\mathbf{H}}_{ir} =
\mathbf{0}$. Any feasible $\mathbf{u}_i$ must lie in the orthogonal
subspace of $\overline{\mathbf{H}}_{ir}$. Therefore, we have
\begin{align}
\label{eq:U_hir_ortho} \mathbf{u}_i =
\mathbf{S}_\bot(\overline{\mathbf{H}}_{ir})\mathbf{x},
\end{align}
where $\mathbf{x}$ is an arbitrary vector. Then the optimization
problem (\ref{eq:heursi_U_sub2}) with fixed
$\overline{\mathbf{U}}_i$ can be rewritten as follows by
substituting (\ref{eq:U_hir_ortho}),
\begin{align}
\label{eq:U_i_x_i} \underset{\mathbf{x}}{\max~} &
\lVert\mathbf{x}^T \mathbf{S}_\bot(\overline{\mathbf{H}}_{ir})^T\mathbf{H}_{br}S_{\bot}(\overline{\mathbf{U}}_i^T\mathbf{H}_{br})\rVert\nonumber\\
\text{s.t.~} & \lVert \mathbf{x}\rVert = 1.
\end{align}
The optimal value of $\mathbf{x}$ is the left singular vector of
$\mathbf{S}_\bot(\overline{\mathbf{H}}_{ir})^T\mathbf{H}_{br}\mathbf{S}_{\bot}(\overline{\mathbf{U}}_i^T\mathbf{H}_{br})$
corresponding to its largest singular value \cite{VanTree}. Then
from (\ref{eq:U_hir_ortho}), we can obtain the optimal
$\mathbf{u}_i$.

Substituting the optimization result $\mathbf{U}^\star$ into
(\ref{eq:Wr_D structure}), the RS weighting matrix designed for
maximizing the downlink sum rate can be obtained as
$\mathbf{W}_r^{\star1}=(\mathbf{H}_{ur}^T)^{\dag}\mathbf{G}_{r1}{\mathbf{U}^\star}^T$.

\subsubsection{Design of $\mathbf{W}_r$ From
Subproblem (\ref{eq:Wr_U problem})} We can also show that the
optimal RS weighting matrix that maximizes the uplink sum rate has
the following structure,
\begin{align}
\label{eq:WrU solution up} \mathbf{W}_r^{\star2} = \mathbf{U}^\star
\mathbf{G}_{r2}\mathbf{H}_{ur}^{\dag},
\end{align}
where $\mathbf{U}^\star$ and $\mathbf{G}_{r2}$ can be obtained
similarly as in the last subsection. We do not present the detailed
derivation for concision.

\begin{table*}[htp]
\renewcommand{\arraystretch}{1.3}
\caption{Comparison of Computational Complexity}
\label{ta:complexity}
\begin{center}
\begin{tabular}{c|p{3.9cm}|c|p{3.5cm}|c}
\hline \hline
          & \multicolumn{2}{|c|}{RS} & \multicolumn{2}{|c}{BS}\\
\hline
          & major operations & complexity & major operations & complexity\\
\hline
ZF scheme        & pseudo inverse of a $N_R\times 2N_U$ matrix & $\mathcal{O}(N_RN_U^{2})$ & none & 0\\
\hline
SA scheme        & pseudo inverse of a $N_R\times N_U$ matrix & $\mathcal{O}(N_RN_U^{2})$   & pseudo inverse of a $N_B\times N_U$ matrix & $\mathcal{O}(N_BN_U^{2})$\\
\hline Balanced scheme & \tabincell{p{3.9cm}}{pseudo inverse of a
$N_R\times N_U$ matrix,\\ $N_U$ times of SVD of $N_R\times
(N_B-N_U+1)$ matrix}
                                                & \tabincell{c}{$\mathcal{O}(N_RN_U^{2}) +$\\ $\mathcal{O}(N_RN_U(N_B-N_U+1)^2)$} & pseudo inverse of $N_B\times N_U$ matrices & $\mathcal{O}(N_BN_U^{2})$\\
\hline \hline
\end{tabular}
\end{center}
\end{table*}

\subsubsection{Balancing $R_U$ and $R_D$ to Maximize
Bidirectional Sum Rate} \label{sec:heuristic_balance} Consider that
bidirectional sum rate $R_S = R_U + R_D$, while
$\mathbf{W}_r^{\star1}$ and $\mathbf{W}_r^{\star2}$ are respectively
optimized for $R_U$ and $R_D$. To improve $R_S$, we propose the
following RS weighting matrix,
\begin{align}
\label{eq:Wr Balance} \mathbf{W}_r^{BL} =
c_\gamma(\gamma\mathbf{W}_r^{\star1} +
(1-\gamma)\mathbf{W}_r^{\star2}),
\end{align}
where the power adjusting factor $\gamma$, $0 \leq \gamma\leq 1$, is
used for controlling the power proportion to $\mathbf{W}_r^{\star1}$
and $\mathbf{W}_r^{\star2}$ to balance the uplink and downlink sum
rate, $c_\gamma$ is used to meet the total RS transmit power
constraint. In practical systems, after obtaining
$\mathbf{W}_r^{\star1}$ and $\mathbf{W}_r^{\star2}$, the RS can
search for an optimal $\gamma$ that maximizes the bidirectional sum
rate.

\emph{Remark 2:} In a TWR system with single user and single-antenna
BS, $\mathbf{W}_r^{BL}$ turns out to be a maximal-ratio combination
and maximal-ratio transmission (MRC-MRT) weighting matrix. It is
shown in \cite{Zhang2009} that the  bidirectional sum rate gap
between MRC-MRT and the optimal scheme is no more than 0.2bps/Hz.
Though in multi-user multi-antenna TWR system, we cannot draw the
same conclusion via rigorously analysis, the simulation results in
section \ref{sec:simulation} will show that such a balanced solution
performs closely to the alternating optimization solution.

\emph{Remark 3:} \label{sec:compare_with_SA} By substituting
$\mathbf{W}_r^{\star1}$ and $\mathbf{W}_r^{\star2}$ into (\ref{eq:Wr
Balance}), we have
\begin{align}
\label{eq:HUvsSA_HU1} \mathbf{W}_r^{BL} =
c_\gamma(\gamma(\mathbf{H}_{ur}^T)^{\dag}\mathbf{G}_{r1}\mathbf{U}^{\star
T} + (1-\gamma)\mathbf{U}^\star
\mathbf{G}_{r2}\mathbf{H}_{ur}^{\dag}).
\end{align}

As shown in (\ref{eq:U_hir_ortho}), each column of matrix
$\mathbf{U}^\star$ lies in the orthogonal subspace of
$\overline{\mathbf{H}}_{ir}$, i.e., $\mathbf{u}_i^{\star}\in
\mathbf{S}_{\bot}(\overline{\mathbf{H}}_{ir})$. The $i$th column of
$(\mathbf{H}_{ur}^T)^\dagger$ also lies in that orthogonal subspace,
i.e., $\mathbf{q}_i\in
\mathbf{S}_{\bot}(\overline{\mathbf{H}}_{ir})$.

When $N_R=N_U$, i.e., the number of RS antennas equals to the number
of users, the matrix $\overline{\mathbf{H}}_{ir}$ is a $N_R\times
(N_R-1)$ matrix. Therefore, the rank of its orthogonal subspace
$\mathbf{S}_{\bot}(\overline{\mathbf{H}}_{ir})$ is one. Since both
of $\mathbf{u}_i^{\star}$ and $\mathbf{q}_i$ lie in the same rank-1
subspace, they are linearly dependent, i.e., $\mathbf{u}_i^{\star} =
d_i\mathbf{q}_i$, where $d_i$ is a scalar. Therefore, we have
$\mathbf{U}^\star = (\mathbf{H}_{ur}^T)^\dagger\mathbf{D}$, where
$\mathbf{D} = \text{diag}(d_1,\cdots,d_{N_U})$. Substituting this
expression into (\ref{eq:HUvsSA_HU1}), we obtain
\begin{align}
\label{eq:HUvsWr HU2} \mathbf{W}_r^{BL} & =
(\mathbf{H}_{ur}^T)^\dagger(c_\gamma\gamma\mathbf{G}_{r1}\mathbf{D}^T
+c_\gamma(1-\gamma)\mathbf{D}\mathbf{G}_{r2})\mathbf{H}_{ur}^\dagger\nonumber\\
& \triangleq
(\mathbf{H}_{ur}^T)^\dagger\mathbf{G}_r^{BL}\mathbf{H}_{ur}^\dagger,
\end{align}
where $\mathbf{G}_r^{BL} \triangleq
c_\gamma\gamma\mathbf{G}_{r1}\mathbf{D}^T
+c_\gamma(1-\gamma)\mathbf{D}\mathbf{G}_{r2}$ is a diagonal matrix.

Comparing (\ref{eq:HUvsWr HU2}) with the RS transceiver in the SA
scheme proposed in \cite{Yang2008a,Toh2009,DK2011}, we see that
$\mathbf{W}_r^{BL}$ has the same form as that of the SA scheme.
Substituting (\ref{eq:HUvsWr HU2}) into (\ref{eq:Wbt_heuristic}) and
(\ref{eq:Wbr_heuristic}), it is easy to show that the BS
transceivers in our balanced solution also have the same forms as
those in the SA scheme. This indicates that the SA scheme is a
special case of the balanced solution when $N_R=N_U$. In fact, in
such a setting, it is not hard to show that the solution of
interference free constraints (\ref{eq:constraint 1 for precoder}),
(\ref{eq:constraint 2 for precoder}) and (\ref{eq:constraint 3 for
precoder}) is unique, which is exactly the SA scheme.

\subsection{Complexity Comparison}
Here we compare the computational complexities of the balanced
scheme and the existing ZF \cite{Esli2008a} and SA schemes
\cite{Toh2009,DK2011,Yang2008a}.

In the ZF scheme, the major operation at the RS is to compute the
pseudo inverse of a $N_R\times 2N_U$ matrix. Since all the
interference are eliminated by the RS, the BS needs to do nothing.
In the SA scheme, the major operations at the RS and the BS are to
compute the pseudo inverses of a $N_R \times N_U$ matrix and a
$N_B\times N_U$ matrix, respectively.

In the balanced scheme, to obtain the RS transceiver (\ref{eq:Wr
Balance}), we need to compute $\mathbf{H}_{ur}^\dagger$,
$\mathbf{U}$, $\mathbf{G}_{r1}$ and $\mathbf{G}_{r2}$, and search
for the optimal balancing factor $\gamma$. Specifically, we need to
perform $N_U$ times of singular vector decomposition (SVD) to
alternately design the $N_U$ columns of $\mathbf{U}$. According to
(\ref{eq:U_i_x_i}), each SVD is performed for a $N_R \times
(N_B-N_U+1)$ matrix. Only vector norm operation is required to
compute the power allocation matrices $\mathbf{G}_{r1}$ and
$\mathbf{G}_{r2}$, for which the complexity can be neglected
compared with those of the pseudo inverse and SVD. The complexity of
finding the optimal $\gamma$ can also be ignored, which only
requires a scalar searching operation. To obtain the BS transceivers
in our balanced scheme (\ref{eq:Wbt_heuristic}) and
(\ref{eq:Wbr_heuristic}), we need to compute the pseudo inverses of
two $N_B\times N_U$ matrices.

A widely used method for computing pseudo inverse is using SVD,
which results in a complexity of $\mathcal{O}(mn^2)$ flops to
compute the pseudo inverse of a $m\times n$ matrix \cite{Xu1994},
where $m\geq n$. The complexities of the transceiver schemes are
present in Table~\ref{ta:complexity}, which shows that the
complexity of the balanced scheme is on the same order as those of
the SA and ZF schemes.

\section{Simulation Results}
\label{sec:simulation} In this section, we evaluate the performance
of the proposed transceivers and compare them with existing schemes
by simulations. We assume that all channels are independent and
identically distributed Rayleigh fading channels, and all simulation
results are obtained by averaging over 1000 Monte-Carlo trails. For
a fair comparison, we use equal power allocation at the BS and RS in
all the transceiver schemes. We assume that the noise variance $N_0$
is identical at the BS, RS and each user. The transmit power of each
user $P_U=1$. The BS and RS transmit power are $P_B$ and $P_R$,
respectively. We define $1/N_0$ as the transmit SNR. Without
otherwise specified, we set $N_B=2$, $N_R=4$, $N_U=2$, $P_B=P_R=2$,
and $SNR=30$ dB.

\subsection{Impact of the Adjusting factor}
\begin{figure}[htp]
\centering
\includegraphics[width=3.4in]{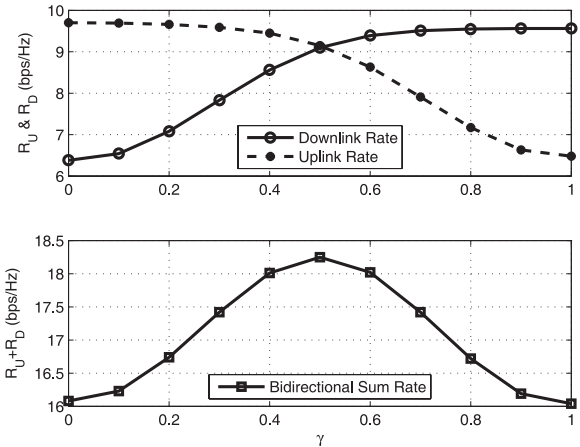}
\caption{Sum rates vs. the power adjusting factor $\gamma$, $N_B$ =
2, $N_R$ = 4, and $N_U$ =2} \label{fig:rateVSgamma}
\end{figure}

The sum rates of balanced scheme versus the power adjusting factor
$\gamma$ are shown in Fig.~\ref{fig:rateVSgamma}, where the upper
sub-figure shows the uplink and downlink sum rates and the lower
sub-figure shows the bidirectional sum rate. When $\gamma = 0$, the
RS weighting matrix $\mathbf{W}_r = \mathbf{W}_r^{\star2}$, which
aims to maximize the uplink sum rate. Therefore, the system achieves
high uplink sum rate but low downlink sum rate in this case. By
contrast, when $\gamma = 1$, the system achieves high downlink rate
but low uplink rate. By adjusting the value of $\gamma$, the uplink
and downlink performance are balanced and higher bidirectional sum
rate is achieved. The optimal $\gamma$ under this case is 0.5.

\subsection{Convergence of the Alternating Optimization Solution}
To study the convergence of the alternating optimization algorithm,
we respectively use the proposed balanced transceiver, the ZF and SA
transceivers and multiple random weighting matrices as its initial
value. When using random matrices as the initial values, we pick one
from multiple results that converges to the highest sum rate.

\begin{figure}[htp]
\centering
\includegraphics[width=3.2in]{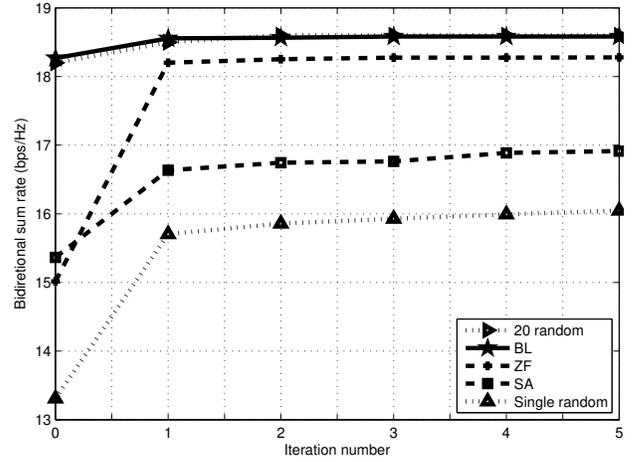}
\caption{Convergence of the alternating optimization algorithm with
different initial values, $N_B$ = 2, $N_R$ = 4, and $N_U$ =2}
\label{fig:sim_convergence}
\end{figure}

Figure~\ref{fig:sim_convergence} shows the bidirectional sum rate
versus the iteration number. The sum rate converges rapidly but the
converged result depends on the initial values due to the
non-convexity nature of the optimization problem. Nonetheless, by
using multiple random initial values, higher bidirectional sum rate
can be achieved. We observe from extensive simulations that when the
number of random initial values exceeds 20, the performance gain is
marginal. Therefore, we can take the result with 20 random initial
values as a near-optimal result. It is shown that the performance of
the balanced transceiver is very close to that of the near-optimal
result. In the following, we will use the balanced transceiver as
the initial value for the alternating optimization.

\subsection{Comparison among Different Transceivers}
We compare the bidirectional sum rates of alternating optimization
solution and the balanced transceiver with those of the ZF
\cite{Esli2008a} and SA schemes \cite{Yang2008a,Toh2009,DK2011}. We
also compare with a minimum-mean-square-error (MMSE) transceiver
without the interference free constraints, where the MMSE BS
transceiver and MMSE RS transceiver were alternately optimized
\cite{Degenhardt2011}.

\begin{figure}[htp]
\centering
\includegraphics[width=3.2in]{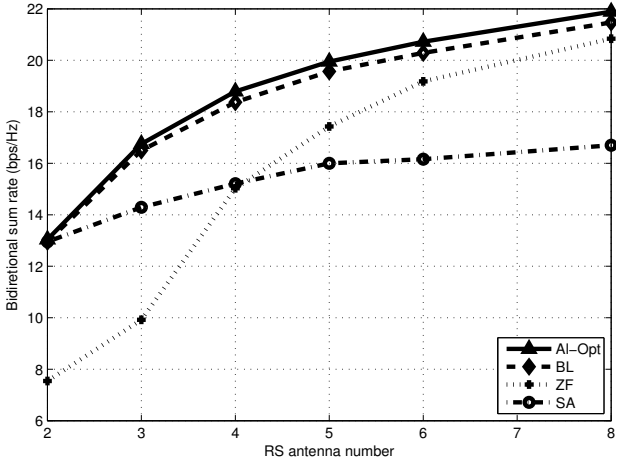}
\caption{Sum rates of four transceivers vs. RS antenna number, $N_B$
= 2, and $N_U$ =2} \label{fig:sim_NR}
\end{figure}

Figure \ref{fig:sim_NR} shows the impact of the antenna number of
the RS, where ``Al-Opt'' denotes the alternating optimization
solution. When there are two users, the ZF scheme needs at least 4
antennas at the RS to cancel all the interference, while the SA
scheme only needs 2 antennas. From the simulation results, we see
that when $N_R \leq 4$ the sum rate of the ZF scheme reduces sharply
due to the residual IUI, but the SA scheme performs much better.
When $N_R > 4$, the ZF scheme becomes superior because it can remove
all IUI but the SA scheme suffers from a power loss when aligning
the downlink signals with the uplink signals. The sum rate of the
balanced transceiver is close to that of the alternating
optimization solution, both are higher than the existing ZF and SA
schemes for any antenna number at the RS.

\begin{figure}[htp]
\centering
\includegraphics[width=3.2in]{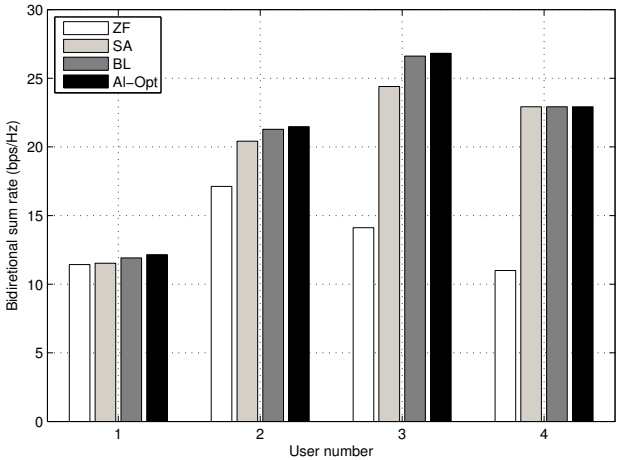}
\caption{Sum rate of four transceivers vs. user number, $N_B$ = 4,
and $N_R$ = 4} \label{fig:sim_NU}
\end{figure}

Figure \ref{fig:sim_NU} shows the impact of the user number on the
performance of different transceivers. We set $N_B=N_R=4$, and
$P_B=P_R=4$. Round robin scheduler is applied, where the scheduled
user number $N_U$ is from 1 to 4. It shows that the performance of
the ZF scheme degrades severely  when $N_U
> 2$ because the four-antenna RS can not cancel all IUI. With the SA
scheme, the proposed balanced scheme and the alternating
optimization solution, the system achieves the highest bidirectional
sum rate when three users are scheduled, where both the balanced
scheme and alternating optimization result have about 2bps/Hz sum
rate gain over the SA scheme. When $N_U=4$, we see that the
performance of the balanced transceiver and the SA scheme are
exactly the same. This agrees well with our earlier analysis in
\emph{Remark 3}.

\begin{figure}[htp]
\centering
\includegraphics[width=3.2in]{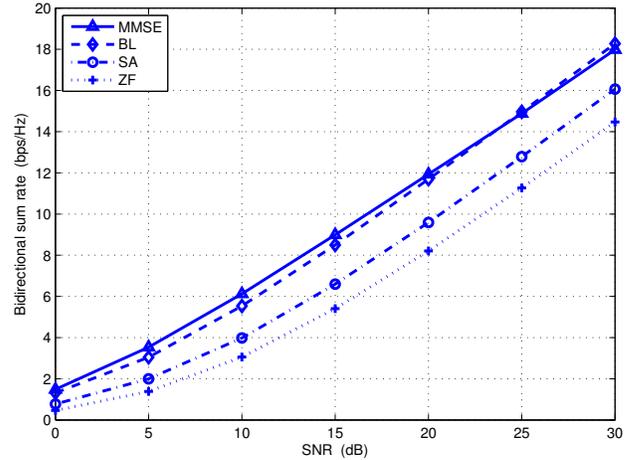}
\caption{Sum rate vs. SNR, $N_B$ = 2, $N_R$ = 4, and $N_U$ = 2.}
\label{fig:MMSE}
\end{figure}

In Fig.~\ref{fig:MMSE} we compare the sum rate of the interference
free transceiver schemes with the MMSE transceiver
\cite{Degenhardt2011}. We can see that our balanced scheme provides
higher sum rate than the existing ZF and SA schemes in a wide range
of transmit SNR. The MMSE scheme is slightly superior to our
balanced scheme in low SNR region, but is inferior to the proposed
scheme in high SNR region. This is because the MMSE solution in
\cite{Degenhardt2011} is obtained via alternating optimization,
which is not guaranteed to be globally optimal. In high SNR region,
the system is interference-limited, therefore the proposed scheme
outperforms the MMSE solution by removing all the interference.

\begin{figure}[htp]
\centering
\includegraphics[width=3.2in]{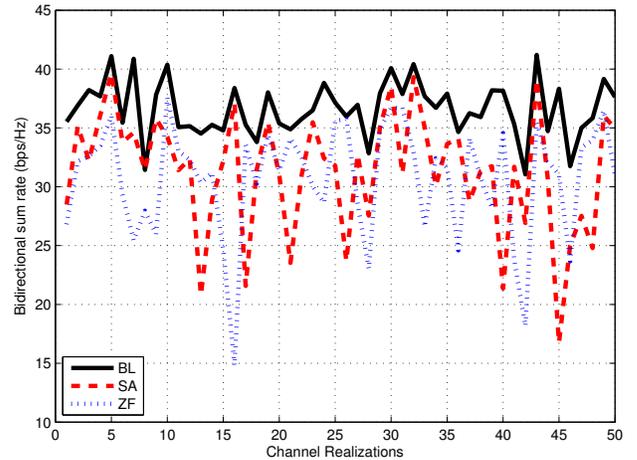}
\caption{Sum rate under different channel realizations, $N_B$ = 2,
$N_R$ = 4, and $N_U$ = 2.} \label{fig:singleCH}
\end{figure}

In Fig.~\ref{fig:singleCH}, we provide the sum rate under each
single channel realization to understand the behavior of the IUI
free transceivers. We see that the ZF and SA schemes perform
differently for a given channel. The ZF scheme requires the RS to
separate all the signals transmitted by the users and BS, and
performs well only when the channel vectors from the users and the
BS are mutually orthogonal. Contrarily, the SA scheme needs to align
the signals transmitted by the BS onto the same directions of the
signals transmitted by the users, and thus performs well only when
the channel vectors from the users and those from the BS have the
same direction. Our balanced scheme can adaptively adjust
transmission strategy depending on the channel condition to ensure
IUI free without the requirements for channel ``orthogonalization''
or ``alignment''. Therefore, its sum rate is always higher than
those of ZF and SA schemes.

\begin{figure}[htp]
\centering
\includegraphics[width=3.2in]{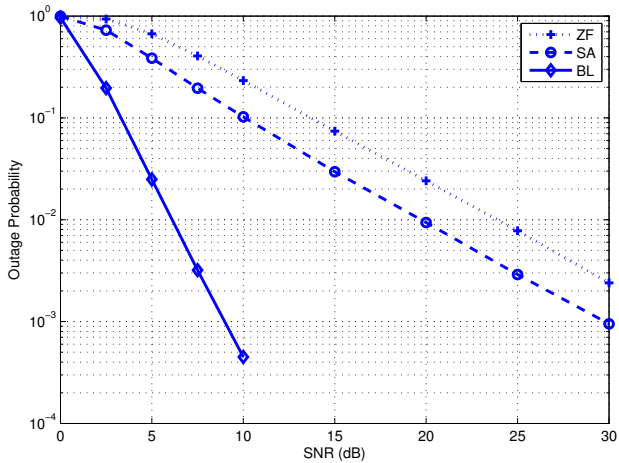}
\caption{Outage probability vs. SNR, $N_B$ = 2, $N_R$ = 4, and $N_U$
= 2.} \label{fig:outage}
\end{figure}

In Fig.~\ref{fig:outage}, we compare the outage probabilities of the
IUI free transceivers with $10^5$ Monte-Carlo trails, where the
system is in outage if its bidirectional sum rate drops below a
given threshold, which is set as 2bps/Hz. We see that our balanced
scheme achieves much lower outage probability than both the ZF and
SA schemes. Moreover, since the sum rate of the balanced scheme is
always ``riding on the peak'' of the ZF and SA schemes as shown in
Fig.~\ref{fig:singleCH}, the balanced scheme achieves higher
diversity gain, and its outage probability decreases much faster
than those of the ZF and SA schemes as the SNR increases.

\section{Conclusion}
\label{sec:conclusion} In this paper, we have designed transceiver
for multi-user multi-antenna two-way relay systems. We first
employed alternating optimization to find the BS and RS transceivers
that maximizes the bidirectional sum rate under interference free
constraints. We proceeded to propose a low complexity balanced
transceiver scheme. By analyzing the solution of the alternating
optimization in high transmit power region, we find that
zero-forcing BS transceivers are asymptotically optimal. Given the
BS transceivers, we designed the RS transceivers by respectively
maximizing the uplink and downlink rate, which are then combined
with a power adjustment factor to maximize the bidirectional sum
rate. Existing signal alignment scheme was shown as a special case
of the balance scheme where the relay antenna number equals to the
user number. Simulation results showed that the performance gap
between the balanced scheme and the alternating optimization
solution is minor. In general system settings, the bidirectional sum
rate of the balanced transceiver is higher than the existing signal
alignment and zero-forcing schemes.

\appendix
\subsection*{Proof of the Optimal Structure of $\mathbf{W}_r$
in (\ref{eq:Wr_D structure})}

Define $\mathbf{V}_{ur} \in \mathbb{C}^{N_R\times N_U}$ as a matrix
consisting of the $N_U$ singular vectors of $\mathbf{H}_{ur}^T$, and
$\mathbf{V}_{ur}^{\perp} \in \mathbb{C}^{N_R\times (N_R-N_U)}$ as a
matrix consisting of the $N_R - N_U$ singular vectors of the
orthogonal subspace of $\mathbf{H}_{ur}^T$. Then
$\mathbf{V}_F=[\mathbf{V}_{ur}~\mathbf{V}_{ur}^{\perp}]$ is a
unitary matrix, and $\mathbf{W}_r$ can be expressed as
\begin{align}
\label{eqA:Wr general} \mathbf{W}_r = \mathbf{V}_F\mathbf{V}_F^H
\mathbf{W}_r =\mathbf{V}_{ur}\mathbf{A}^T +
\mathbf{V}_{ur}^{\perp}\mathbf{B}^T,
\end{align}
where $\mathbf{A} \in \mathbb{C}^{N_R\times N_U}$, $\mathbf{B} \in
\mathbb{C}^{N_R\times (N_R-N_U)}$ are two arbitrary matrices.

Since $\mathbf{h}_{ir}^T\mathbf{V}_{ur}^{\perp} = \mathbf{0}$, from
(\ref{eq:SINR}), the downlink sum rate $R_D$ can be written as
\begin{align}
\label{eqA:Rd_notB} R_D & =
\frac{1}{2}\sum_{i=1}^{N_U}\log_2(1+\frac{\lvert\mathbf{h}_{ir}^T\mathbf{V}_{ur}\mathbf{A}^T\mathbf{H}_{br}\mathbf{w}_{bti}\rvert^2}
{N_0\lVert\mathbf{h}_{ir}^T\mathbf{V}_{ur}\mathbf{A}^T\rVert^2+N_0}).
\end{align}

Substituting (\ref{eqA:Wr general}) into (\ref{eq:Wbt_heuristic}),
we have $\mathbf{W}_{bt} =
(\mathbf{H}_{ur}^T\mathbf{V}_{ur}\mathbf{A}^T\mathbf{H}_{br})^{\dag}\mathbf{G}_b$,
which is not a function of $\mathbf{B}$. Therefore, the value of
$\mathbf{B}$ does not affect the constraints
(\ref{eq:Wbt_heuristic}) and (\ref{eq:equal power BS}) in problem
(\ref{eq:Wr_D problem}). According to (\ref{eqA:Rd_notB}), the
objective function $R_D$ of problem (\ref{eq:Wr_D problem}) also
does not depend on $\mathbf{B}$.

Substituting (\ref{eqA:Wr general}) into (\ref{eq:constraint 3 for
precoder}), we obtain $\mathbf{h}_{ir}^T\mathbf{W}_r\mathbf{h}_{jr}
= \mathbf{h}_{ir}^T\mathbf{V}_{ur}\mathbf{A}^T\mathbf{h}_{jr} = 0, i
\not=j$, which shows that the value of $\mathbf{B}$ does not affect
the constraint (\ref{eq:constraint 3 for precoder}) either.

We can show that the RS transmit power is minimized when $\mathbf{B}
= \mathbf{0}$ as follows,
\begin{align*}
&\lVert\mathbf{W}_r\mathbf{H}_{br}\mathbf{W}_{bt}\rVert^2 +
N_0\lVert\mathbf{W}_r\rVert^2\cr
=&\lVert\mathbf{V}_{ur}\mathbf{A}^T\mathbf{H}_{br}\mathbf{W}_{bt}\rVert^2
+\lVert\mathbf{V}_{ur}^{\perp}\mathbf{B}^T\mathbf{H}_{br}\mathbf{W}_{bt}\rVert^2\nonumber\\
&+N_0\lVert\mathbf{V}_{ur}\mathbf{A}^T\rVert^2 +
N_0\lVert\mathbf{V}_{ur}^{\perp}\mathbf{B}^T\rVert^2\cr
\geq&P_B\lVert\mathbf{V}_{ur}\mathbf{A}^T\mathbf{H}_{br}\mathbf{W}_{bt}\rVert^2
+N_0\lVert\mathbf{V}_{ur}\mathbf{A}^T\rVert^2.
\end{align*}

It indicates that for any given $\mathbf{W}_r
=\mathbf{V}_{ur}\mathbf{A}^T + \mathbf{V}_{ur}^{\perp}\mathbf{B}^T$,
we can always find a $\mathbf{W}_r^{\star}
=\mathbf{V}_{ur}\mathbf{A}^T$, which achieves the same downlink rate
$R_D$ as that with $\mathbf{W}_r$ but consumes less RS power.
Therefore, the optimal $\mathbf{W}_r$ for (\ref{eq:Wr_D problem})
should has the structure of $\mathbf{W}_r =
\mathbf{V}_{ur}\mathbf{A}^T$.

Denote the singular value decomposition of $\mathbf{H}_{ur}^T$ as
$\mathbf{U}_{ur}\mathbf{D}_{ur}\mathbf{V}_{ur}^H$, where
$\mathbf{D}_{ur},\mathbf{U}_{ur}$ are both non-singular matrix, then
we have
\begin{align*}
\mathbf{W}_r  & = \mathbf{V}_{ur}\mathbf{A}^T =
\mathbf{V}_{ur}(\mathbf{D}_{ur}^{-1}\mathbf{U}_{ur}^H
\mathbf{U}_{ur}\mathbf{D}_{ur})\mathbf{A}^T\nonumber\\ &=
(\mathbf{H}_{ur}^T)^{\dag}\mathbf{U}_{ur}\mathbf{D}_{ur}\mathbf{A}^T
\triangleq (\mathbf{H}_{ur}^T)^{\dag}\mathbf{M}^T,
\end{align*}
where $\mathbf{M}^T \triangleq
\mathbf{U}_{ur}\mathbf{D}_{ur}\mathbf{A}^T$. Divide the matrix
$\mathbf{M} = (\mathbf{m}_1,\cdots,\mathbf{m}_{N_U})$ into two
matrices, $\mathbf{U} = (\mathbf{u}_1,\cdots,\mathbf{u}_{N_U})$ and
$\mathbf{G}_{r1} = \text{diag}(p_{r1},\cdots,p_{rN_U})$, where
$\mathbf{u}_j \triangleq \mathbf{m}_j/\lVert \mathbf{m}_j \rVert$
and $p_{rj} \triangleq \lVert \mathbf{m}_j \rVert$. Finally, we have
$\mathbf{W}_r =
(\mathbf{H}_{ur}^T)^{\dag}\mathbf{G}_{r1}\mathbf{U}^T$.

\section*{Acknowledgement}
We would like to thank Prof. Zhiquan Luo for the helpful discussions
on optimization techniques, and thank Dr. Tingting Liu for the
helpful discussions on signal alignment.

\bibliographystyle{IEEEtran}
\bibliography{IEEEabrv,MUTWRS}

\begin{biography}[{\includegraphics[width=1in,height=1.25in,clip,keepaspectratio]{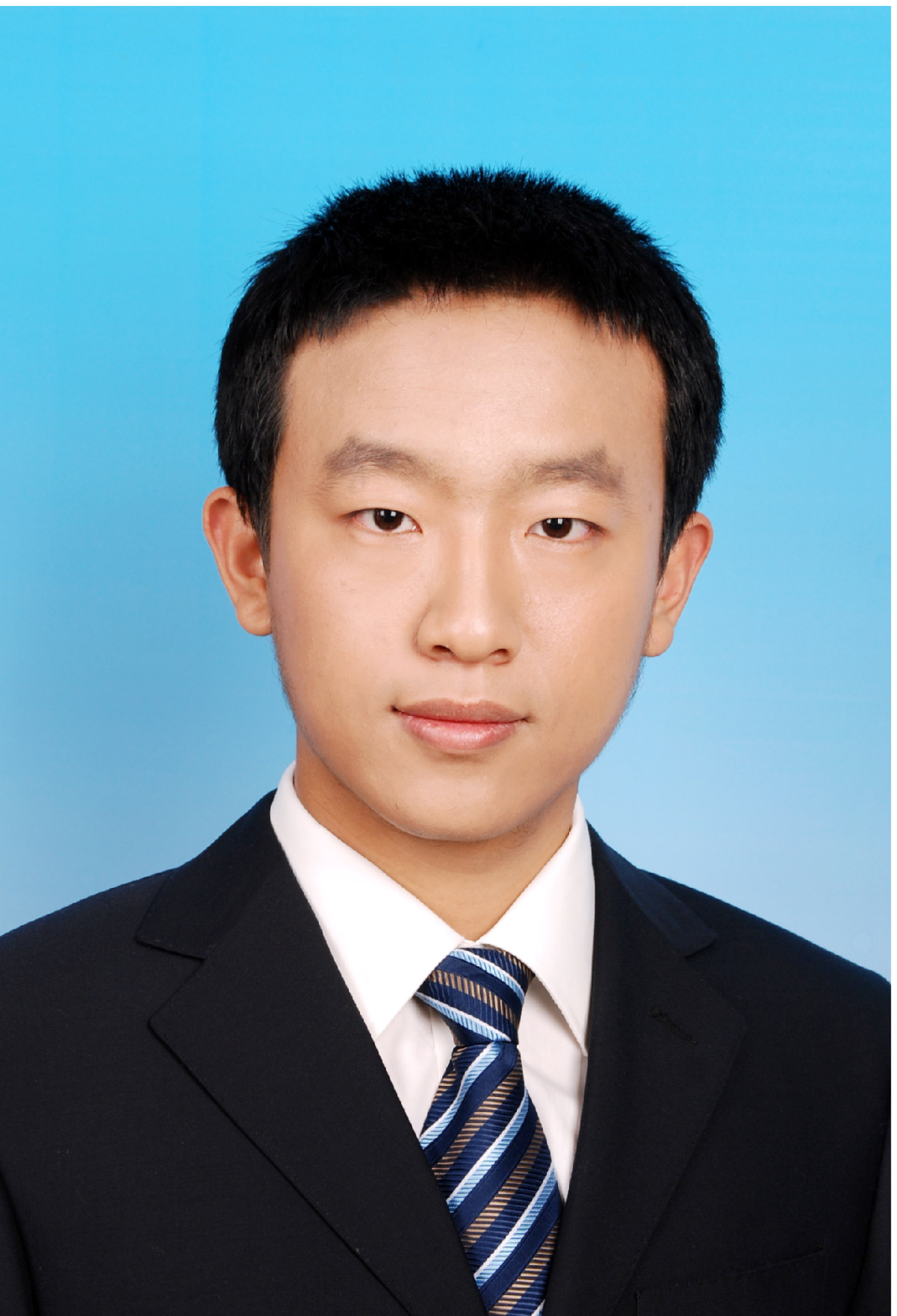}}]{Can Sun}
received his B.S. degree in 2006 from Beijing University of
Aeronautics and Astronautics (BUAA, now renamed as Beihang
University). He is currently a Ph.D student in signal and
information processing in the School of Electronics and Information
Engineering, BUAA, Beijing, China. Since Apr. 2009 to Mar. 2010, he
was a visiting student with the University of Sydney, NSW,
Australia. His research interests include coordinated multi-point
transmission, relay communication and energy efficient transmission.
\end{biography}

\begin{biography}[{\includegraphics[width=1in,height=1.25in,clip,keepaspectratio]{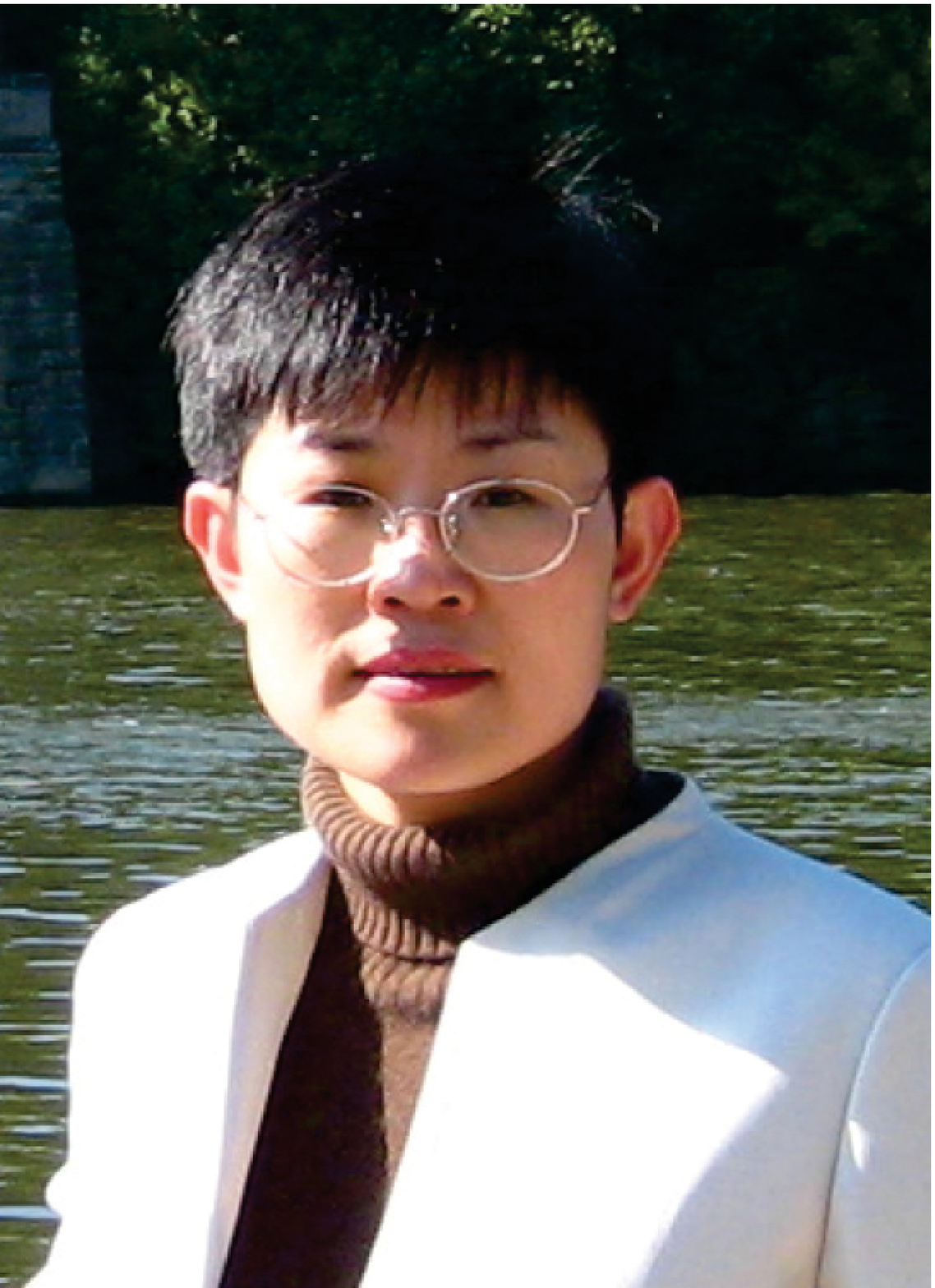}}]{Chenyang Yang}
received her MSE and PhD degrees in 1989 and 1997 in Electrical
Engineering, from Beijing University of Aeronautics and Astronautics
(BUAA, now renamed as Beihang University). She is now a full
professor in the School of Electronics and Information Engineering,
BUAA. She has published various papers and filed many patents in the
fields of signal processing and wireless communications. She was
nominated as an Outstanding Young Professor of Beijing in 1995 and
was supported by the 1st Teaching and Research Award Program for
Outstanding Young Teachers of Higher Education Institutions by
Ministry of Education (P.R.C. "TRAPOYT") during 1999-2004.
Currently, she serves as an associate editor for IEEE Transactions
on Wireless Communications, an associate editor-in-chief of Chinese
Journal of Communications and an associate editor-in-chief of
Chinese Journal of Signal Processing. She is the chair of Beijing
chapter of IEEE Communications Society. She has ever served as TPC
members for many IEEE conferences such as ICC and GLOBECOM. Her
recent research interests include network MIMO, energy efficient
transmission and interference management in multi-cell systems.
\end{biography}

\begin{biography}[{\includegraphics[width=1in,height=1.25in,clip,keepaspectratio]{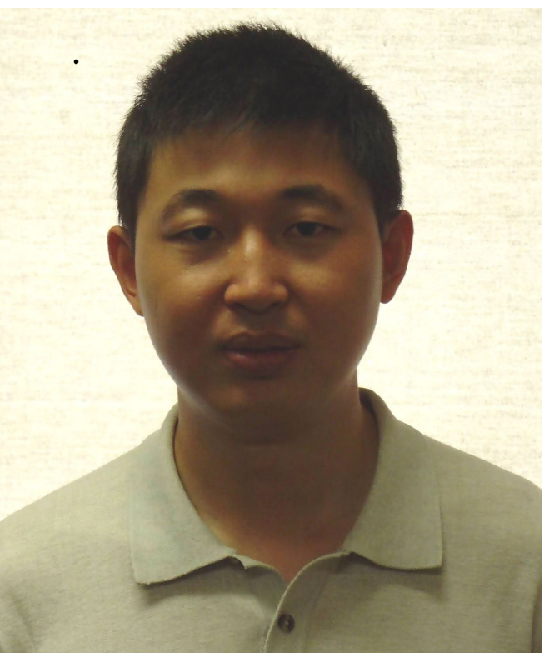}}]{Yonghui Li}
(M'04-SM'09) received his PhD degree in November 2002 from Beijing
University of Aeronautics and Astronautics. From 1999 - 2003, he was
affiliated with Linkair Communication Inc, where he held a position
of project manager with responsibility for the design of physical
layer solutions for the LAS-CDMA system. Since 2003, he has been
with the Centre of Excellence in Telecommunications, the University
of Sydney, Australia. He is now an Associate Professor in School of
Electrical and Information Engineering, University of Sydney. He is
also currently the Australian Queen Elizabeth II fellow. \\
His current research interests are in the area of wireless
communications, with a particular focus on MIMO, cooperative
communications, coding techniques and wireless sensor networks. He
holds a number of patents granted and pending in these fields. He is
an executive editor for European Transactions on Telecommunications
(ETT), Editor for Journal of Networks, and was an Associate Editor
for EURASIP Journal on Wireless Communications and Networking from
2006-2008. He also served as the Leading Editor for special issue on
"advances in error control coding techniques" in EURASIP Journal on
Wireless Communications and Networking, He has also been involved in
the technical committee of several international conferences, such
as ICC, Globecom, etc.
\end{biography}

\begin{biography}[{\includegraphics[width=1in,height=1.25in,clip,keepaspectratio]{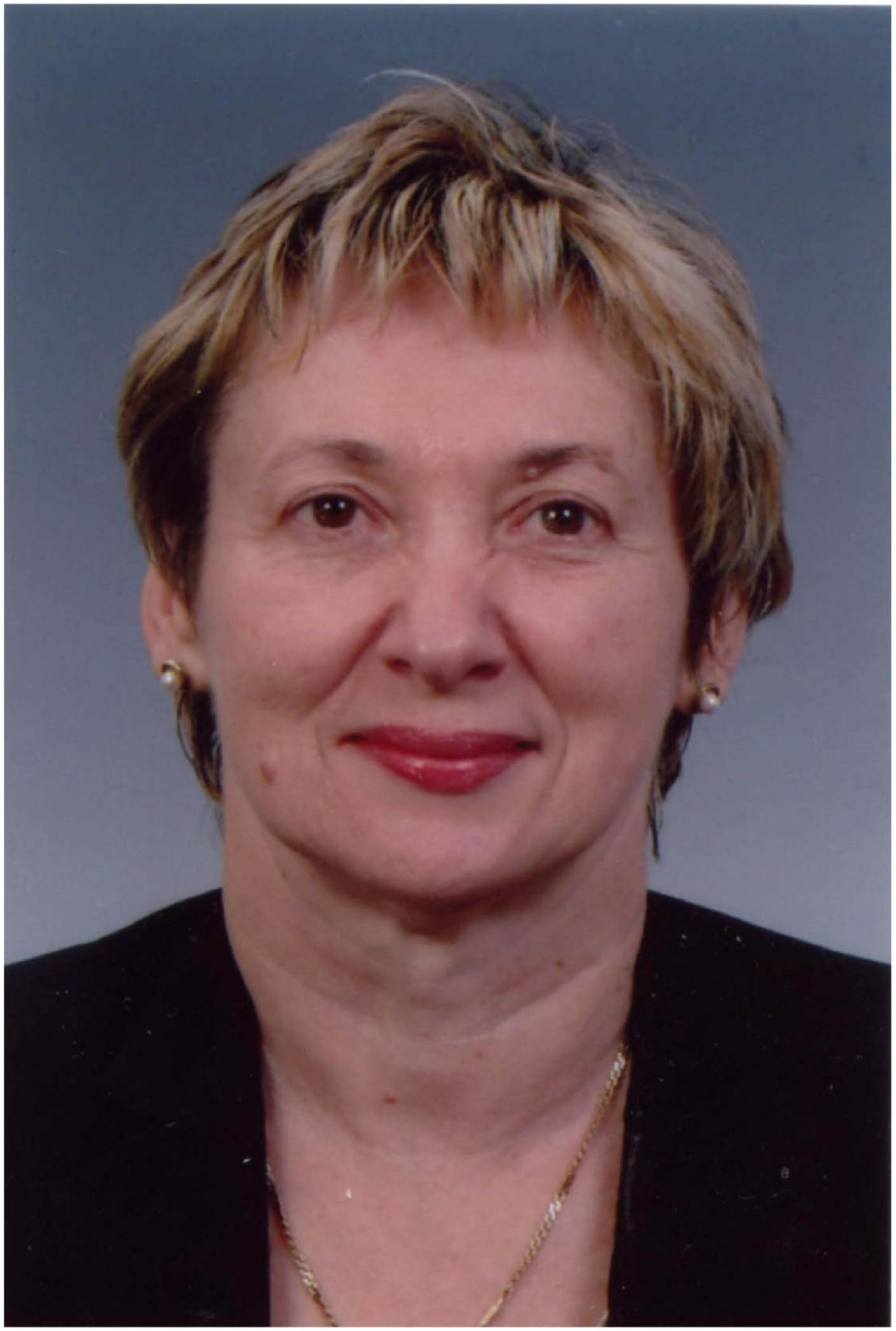}}]{Branka Vucetic}
(M'83-SM'00-F'03) received the B.S.E.E., M.S.E.E., and Ph.D. degrees
in 1972, 1978, and 1982, respectively, in electrical engineering,
from The University of Belgrade, Belgrade, Yugoslavia. During her
career she has held various research and academic positions in
Yugoslavia, Australia, and the UK. Since 1986, she has been with the
Sydney University School of Electrical and Information Engineering
in Sydney, Australia. She is currently the Director of Centre of
Excellence in Telecommunications at Sydney University. Her research
interests include wireless communications, digital communication
theory, coding, and multi-user detection. \\
In the past decade she has been working on a number of industry
sponsored projects in wireless communications and mobile Internet.
She has taught a wide range of undergraduate, postgraduate, and
continuing education courses worldwide. \\
Prof. Vucetic co-authored four books and more than two hundred
papers in telecommunications journals and conference proceedings.
\end{biography}

\end{document}